\pdfoutput=1 
\documentclass[a4paper,11pt]{scrartcl}

\usepackage{subfigure}
\usepackage{hepnames}
\usepackage[load-configurations=abbreviations,tight-spacing=true,separate-uncertainty,bracket-numbers = false]{siunitx}
\usepackage{hyperref}

\usepackage{graphicx} 

\usepackage{authblk}

\newcommand{\eV}{\ensuremath{\mathnormal{eV}}}
\newcommand{\cspeed}{\ensuremath{\mathnormal{c}}}

\newcommand{\Datura}{\ensuremath{\textrm{DATURA}}}

\newcommand{\Mimosa}{\ensuremath{\textrm{MIMOSA\,26}}}

\newcommand{\thetaalu}{\ensuremath{\theta_{\textrm{Al}}}}
\newcommand{\thetameas}{\ensuremath{\theta_{\textrm{meas}}}}
\newcommand{\thetaair}{\ensuremath{\theta_{\textrm{meas,air}}}}

\newcommand{\EUTelescope}{\ensuremath{\textrm{EUTelescope}}}

\begin{document}
\title{Scattering studies with the DATURA beam telescope}
\author[1]{Hendrik Jansen\thanks{corresponding author: hendrik.jansen@desy.de}}
\author[1]{Jan Dreyling-Eschweiler}
\author[1]{Paul Sch\"utze}
\author[2]{Simon Spannagel}
\affil[1]{Deutsches Elektronen-Synchrotron DESY, Hamburg, Germany}
\affil[2]{CERN, Geneva, Switzerland}
\date{}
\maketitle

\flushbottom

\begin{abstract}
High-precision particle tracking devices allow for two-dimensional analyses of the material budget distribution of particle detectors and their periphery.
In this contribution, the material budget of different targets is reconstructed from the width of the angular distribution of scattered beam particle at a sample under test. 
Electrons in the GeV-range serve as beam particles carrying enough momentum to traverse few millimetre thick targets whilst offering sufficient deflection for precise measurement. 
Reference measurements of the scattering angle distribution of targets of known thicknesses are presented that serve as calibration techniques required for tomographic reconstructions
 of inhomogeneous objects. 
\end{abstract}


\section{Introduction}
\label{sec:intro}

Understanding the scattering of charged particles off nuclei in different materials has been of interest for many decades. 
Moli\`ere~\cite{moliere} postulated a theory without empirical parameters to describe multiple scattering in arbitrary materials.
Later, Gaussian approximations to the involved calculations of his theory have been developed e.g.\ by Highland~\cite{ref:scatteringhighland} in order to simplify predictions.

Today, precise tracking detectors allow for the characterisation of unknown materials based on their scattering properties. 
In this contribution, measurements with the $\Datura$ beam telescope, a high-precision tracking device consisting of silicon pixel sensors, are described.
The scattering behaviour of GeV electrons traversing aluminium targets with precisely known thicknesses between \SI{13}{\um} and \SI{e4}{\um} at the DESY test beam facility are studied.
A track reconstruction is performed, enabling the extraction of the particle scattering angles at the target arising from the multiple scattering therein.

\section{Experimental set-up}

The DATURA beam telescope~\cite{pub-datura-paper} consists of six Mimosa26~\cite{ref:mimosa26} monolithic active pixel sensors (MAPS),
 a so-called trigger logic unit (TLU)~\cite{EUDET-2009-04}, four scintillators for triggering purposes, as well as additional infrastructure such as moving stages,
 and the EUDAQ data acquisition system~\cite{ref:tipp2014_eudaq,ref:eudaqwebsite}. 
The $\Mimosa$ sensors feature a pitch of \SI{18.4 x 18.4}{\um} and are thinned down to a thickness of about \SI{50}{\um}.
Together with \SI{50}{\um} thin protective Kapton foil, the total material of the six telescope planes amounts to $\varepsilon = x/X_0 = \num{4.8e-3}$,
 expressed as the fraction of radiation lengths.

\begin{figure}[tb]
  \centering
  \includegraphics[trim = 0 220 250 20, width=0.8\textwidth]{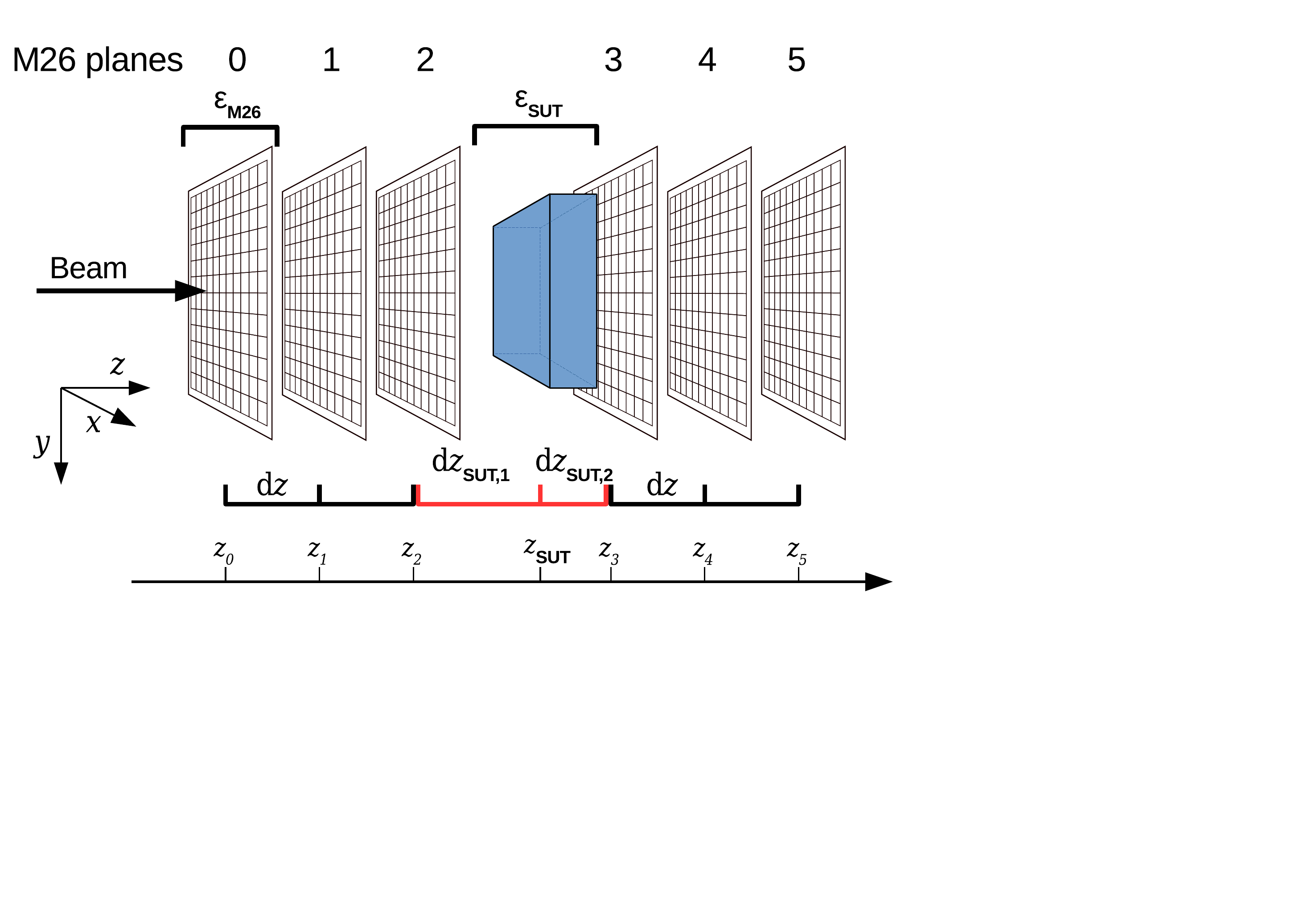}\put(-120,-20){\tiny\copyright H. Jansen 2017, CC-BY 4.0 license}
  \caption[Set-up]{The measurement set-up with its important parameters.}
  \label{fig:setup}
\end{figure}

For the measurements presented, the telescope is operated with equidistant planes, each \SI{20}{\mm} apart, see Figure~\ref{fig:setup}. 
The distance between the last plane of the upstream telescope arm and the upstream surface of sample under test (SUT) amounts to \SI{12.5}{\mm},
 and from the downstream surface of the SUT to the first downstream plane \SI{2.5}{\mm}. 
The intrinsic resolution of the sensors building the DATURA beam telescope allows for precise tracking of beam particles and thus measurements of both the track position
 and the kink angle in the scattering material.
This renders a detailed study of the material budget as a function of the track impact position, and therefore resolved material budget measurements, possible.

In the following, measurements of different targets are presented. 
The measurements have been performed using positrons with energies between \SI{1}{\GeV} and \SI{5}{\GeV} provided by the DESY-II accelerator~\cite{DESYtb}.
The data recorded are analysed using the $\EUTelescope$ software~\cite{EUDET-2008-48,ref:eutelwebsite}. 

\section{Multiple scattering studies}

The width of the angular distribution predicted by Highland's approximation of the Moli\`ere theory for single scatterers evaluates to~\cite{pdg}

\begin{equation}
\label{eq:ms}
  \Theta_0 = \left(\frac{\SI{13.6}{\MeV}}{\beta c p } \cdot z \right) \cdot \sqrt{\varepsilon} \cdot \left( 1 + 0.038 \ln{\varepsilon}\right)
\end{equation}
where $p$, $\beta c$ and $z$ are the momentum, velocity and charge number of the incident particle.
For a composite scatterer, the individual contributions to the material budgets are summed linearly representing the total material budget
%
 $ \varepsilon = \sum_i \varepsilon_i$.
%
The width induced by the $i$-th scatterer therefore reads

\begin{equation}
\label{eq:ms_tele}
  \Theta_{0,i} = \frac{\varepsilon_i}{\varepsilon} \Theta_0 = \left(\frac{\SI{13.6}{\MeV}}{\beta c p } \cdot z \right) \cdot \sqrt{\varepsilon_i}
   \cdot \left( 1 + 0.038 \ln{\varepsilon}\right).
\end{equation}

\noindent
with the correction term still containing the full material budget and not the fraction represented by the individual scatterer.

\subsection{Track model}

The six hits, one in each of the sensor planes, that stem from the same traversing particle are identified using the triplet method,
 which is described in detail in reference~\cite{pub-datura-paper}. 
The General Broken Lines track model~\cite{Blobel2006,Kleinwort2012107} is used to describe the particle trajectories based on the six-tuple found by the triplet method. 
This model includes the notion of scatterers and thus allows for kinks in the trajectory at these positions.
The model neglects  bremsstrahlung effects, non-Gaussian distributed tails as well as non-linear effects
 and describes most accurately the case of the narrow scattering angle approximation. 

 \begin{figure}[tb]
  \centering
  \includegraphics[trim = 0 150 100 50, width=0.8\textwidth]{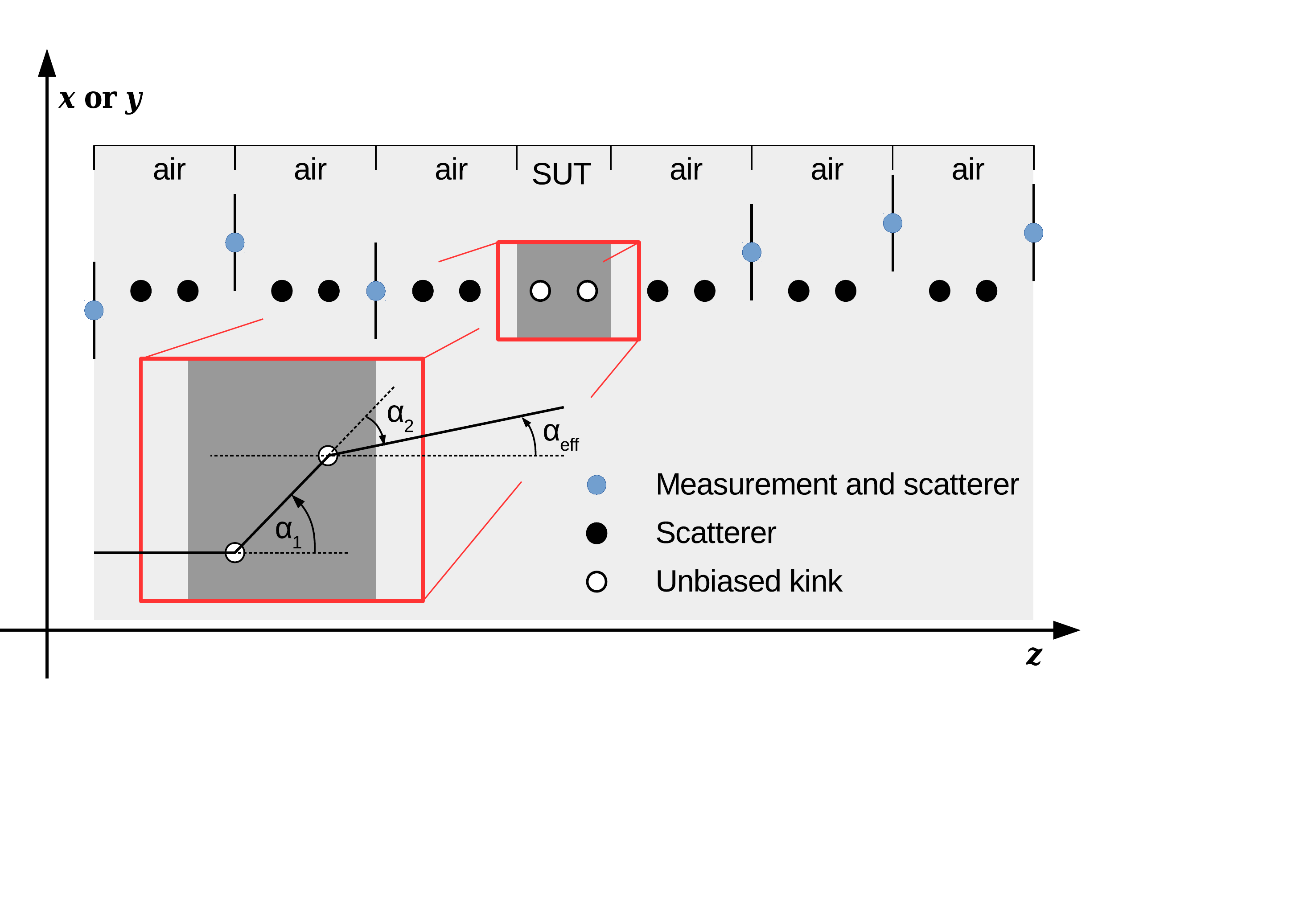}\put(-120,-20){\tiny\copyright H. Jansen 2017, CC-BY 4.0 license}
  \caption[gbl]{The GBL track model with two unbiased kinks at the SUT.}
  \label{fig:gbl}
\end{figure}

For the determination of the material budget of an unknown target new parameters are introduced in the track model. 
For the studies presented, two local derivatives are included at the measurements behind the SUT, which, when appropriately scaled,
 reflect two unbiased kinks at the position of the scattering target, as is shown in Fig.~\ref{fig:gbl}.
This yields an unbiased value for the kink angle at the SUT along two directions, which are chosen along $x$ and $y$. 

\subsection{Homogeneous targets}

\begin{figure}[tb]
  \centering
  \includegraphics[width=0.49\textwidth]{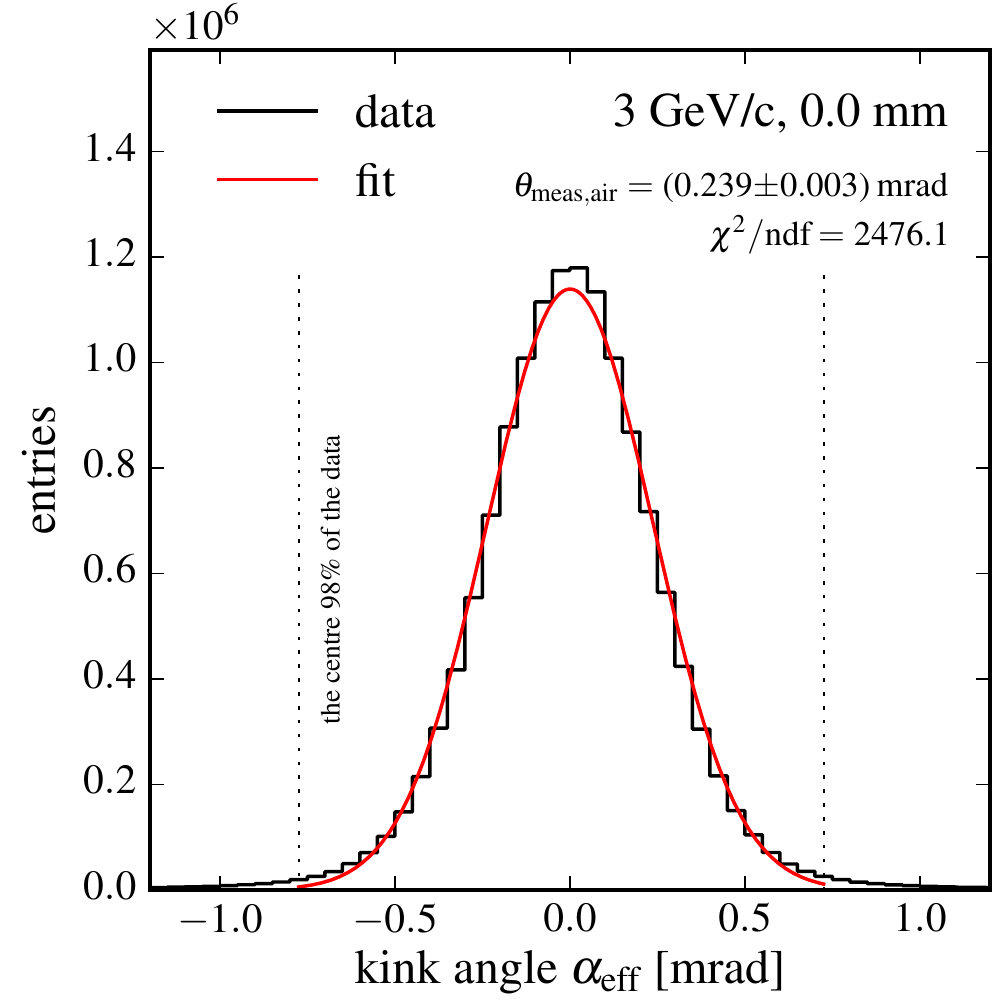}  \put(-190, 4){(A)}
  \includegraphics[width=0.49\textwidth]{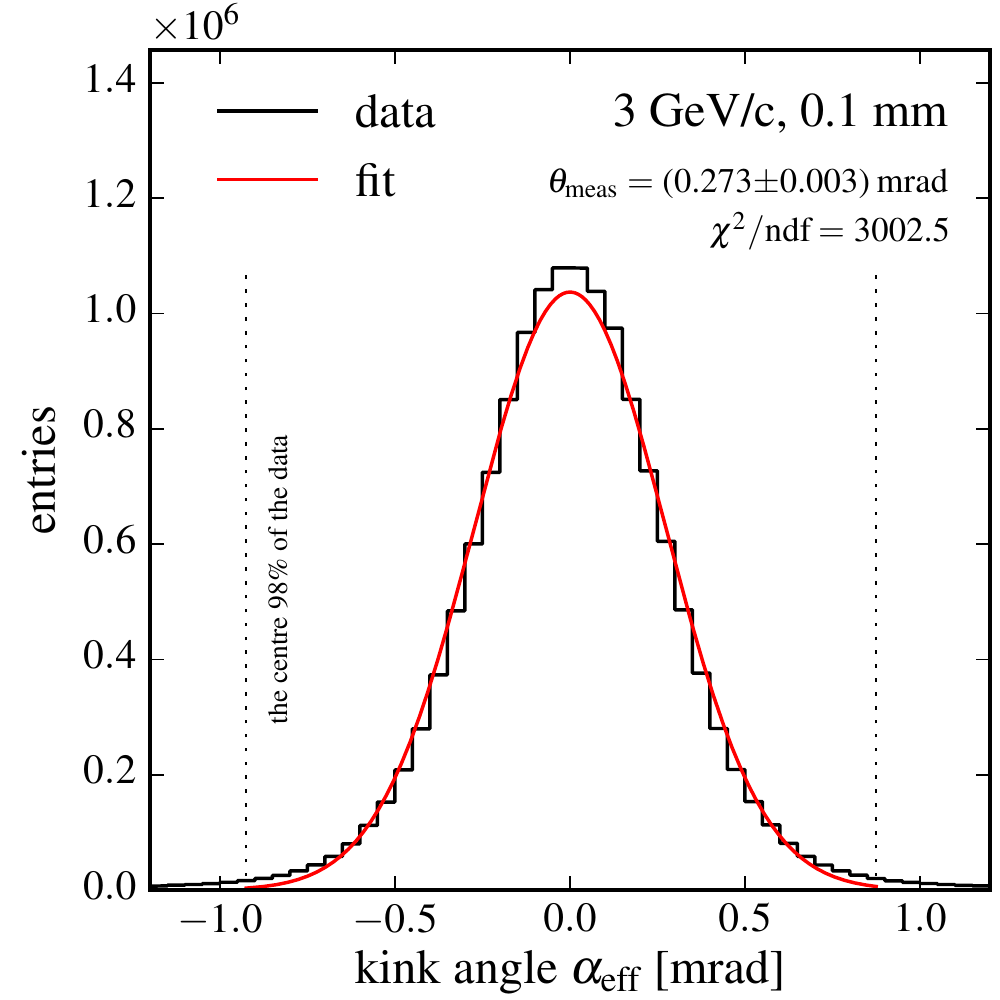} \put(-190, 4){(B)}\put(-120,-20){\tiny\copyright H. Jansen 2017, CC-BY 4.0 license}\\
    \caption[Kink angle distribution]{%
    The kink angle distribution measured at $\SI{3}{\giga\eV/\cspeed}$ for (A) only air and (B) a 0.1\,mm thick aluminium target. 
    A normal distribution is fitted to the centre 98\% of the data yielding the width $\thetameas$ of the measured angle distribution. 
    }
  \label{fig:kinkexample}
\end{figure}

Different homogeneous aluminium targets with thicknesses of 13, 25, 50, 100, 200, 1000 and \SI{e4}{\um} have been measured at various energies
 between $\SI{1}{\giga\eV/\cspeed}$ and $\SI{5}{\giga\eV/\cspeed}$
In addition, a measurement without scattering target has been performed in order to subtract the impact from scattering in air from the results.
Measurements are performed for different beam energies, and the kink angle distributions for different material thicknesses and particle energies are produced with high statistics. 
An example at \SI{3}{\GeV/\cspeed} is shown in Fig.~\ref{fig:kinkexample} for (A) air and (B) \SI{100}{\um} aluminium. 
The distribution is symmetric, centred around zero, but shows clear deviations from a normal distribution. 

\begin{figure}[t]
  \centering
    \includegraphics[width=0.99\textwidth]{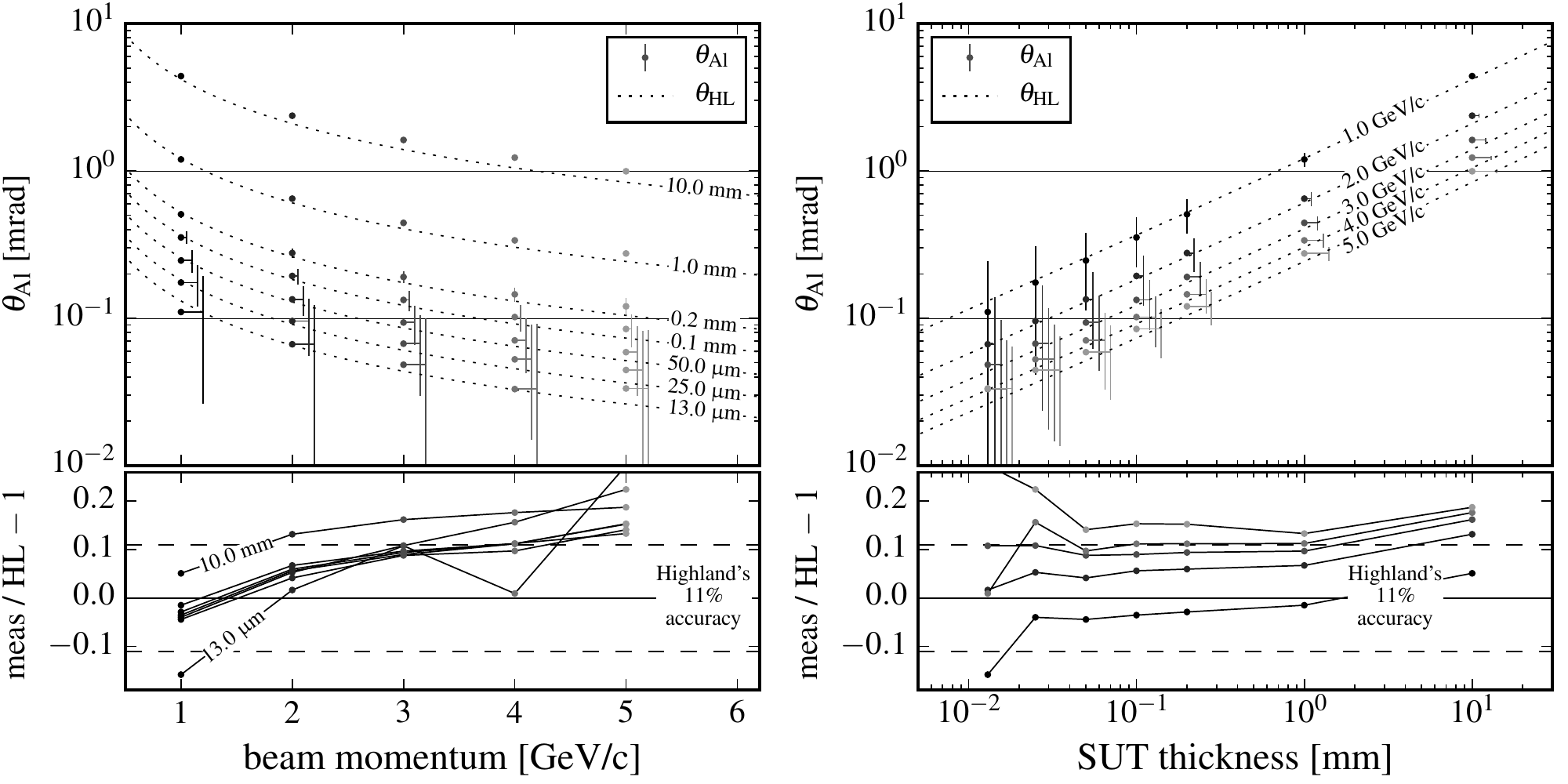}\put(-387, 5){(A)} \put(-175, 5){(B)}\put(-120,-20){\tiny\copyright H. Jansen 2017, CC-BY 4.0 license}
  \caption[Energy and target thickness dependence]{%
  $\thetaalu$ as (A) a function of beam momentum and (B) a function of the target thickness together with Highland prediction.
  The lower plots show the relative deviation between our measurement and the prediction.
  }
  \label{fig:width}
\end{figure}

Figure~\ref{fig:width} presents the width of the kink angle distributions for the different material thicknesses and particle energies.
All measurements are corrected for air by quadratically subtracting the measurement performed without scattering target for the respective energy:
$\thetaalu = \sqrt{\thetameas^2 - \thetaair^2}$. 
In this first analysis a constant systematic uncertainty of 3\,\% is estimated on the values of $\thetameas$ and $\thetaair$ and propagated to $\thetaalu$. 
Figure~\ref{fig:width}~(A) shows $\thetaalu$ as a function of the energy and we observe a monotonically decreasing dependence. 
The data points are presented together with the Highland predictions (dashed) assuming a literature value of $X_0(\textrm{Al}) = \SI{88.97}{\mm}$.
In Fig.~\ref{fig:width}~(B), $\thetaalu$ is plotted as a function of the SUT thickness, following a linear dependence on a double-logarithmic scale. 
The ratio plots display the relative deviation from the Highland prediction, cf.~eq.~(\ref{eq:ms_tele}). 
For the material budget range and the energy range studied, most of the data points lie within a 11\,\% margin~\cite{pdg}, with the thinnest target coming with a sizeable uncertainty. 
Note, that the shape of each distribution depends on the particle energy and the amount of material budget, and hence differ more or less from a normal distribution. 
The systematic dependence of the measurements on the particle energy is under investigation.

\begin{figure}[t!]
  \centering
  \includegraphics[trim = 0 0 0 0, width=0.95\textwidth]{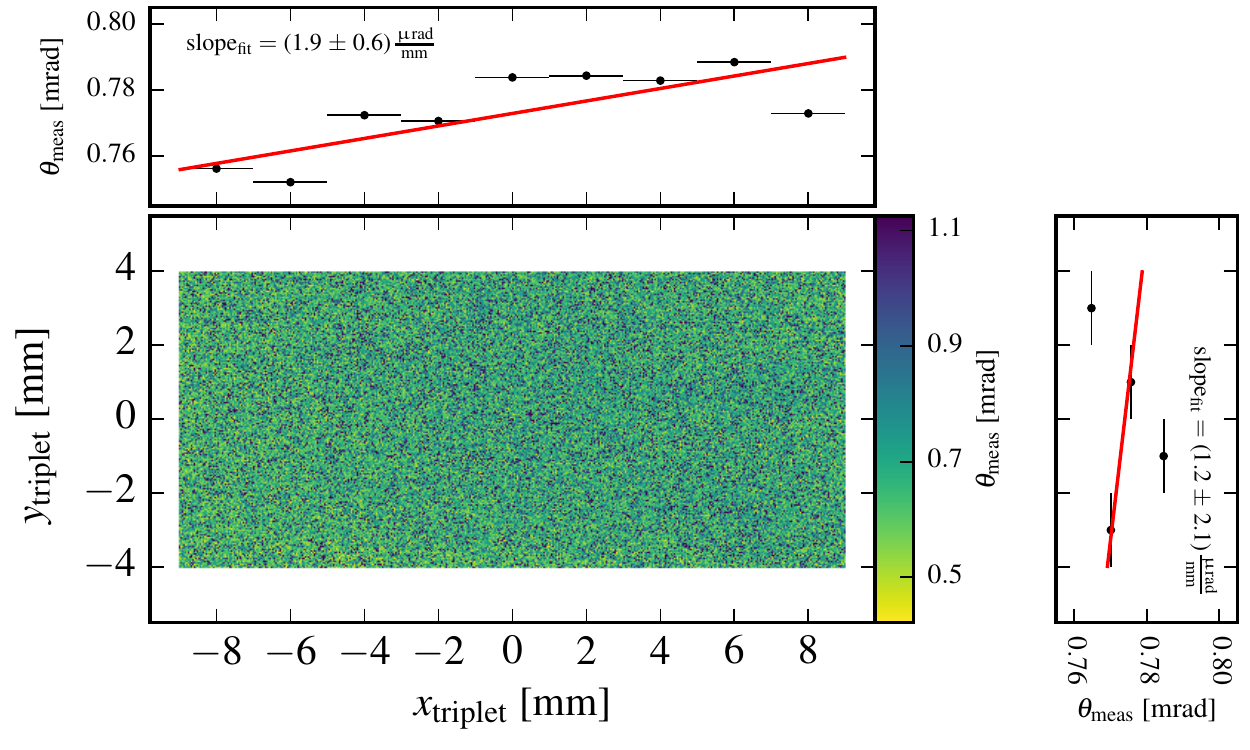} \put(-415, 230){(A)}\\%
  \includegraphics[trim = 0 20 0 0, width=0.95\textwidth]{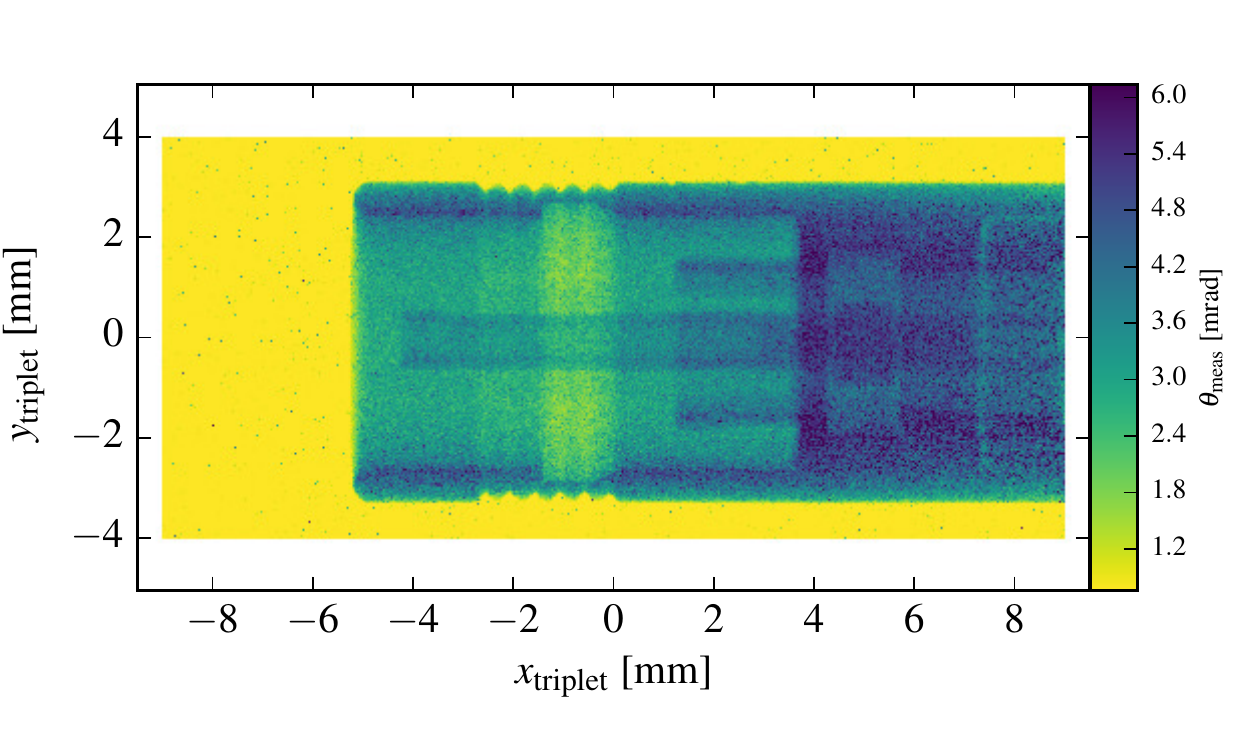} \put(-415, 180){(B)}\put(-120,-20){\tiny\copyright H. Jansen 2017, CC-BY 4.0 license}
    \caption[Position dependence]{%
    (A) The 2D distribution of the kink angle widths and the binned projections in $x$- and $y$-direction
     for an aluminium target a 0.1~mm thickness at $\SI{1}{\giga\eV/\cspeed}$ beam momentum.
    (B) As in (A) for a coaxial connector.}
  \label{fig:means}
\end{figure}

Using the position information of the reconstructed telescope tracks, the width of the kink angle distributions is calculated
 as a function of the position on the scattering target, cf.~Fig.~\ref{fig:means}~(A).
Each $40\times\SI{40}{\um^2}$ cell contains ($80\pm20$) tracks and the corresponding width is the standard deviation of the mean.
We observe a positive trend over the horizontal axis.
Most likely this is an artefact of the generation of the test beam from the DESY-II primary beam.
A carbon fibre target is used to produce bremsstrahlung photons from the primary particles, which undergo pair production on a copper target.
The final momentum selection is performed by a dipole magnet in the horizontal direction and a beam collimator.
Thus, a slight energy dependence in the horizontal axis is expected.
A detailed simulation and a corresponding correction are under study.

\subsection{Inhomogeneous targets}
%

The excellent position resolution provided by the beam telescope renders the measurement of the material budget distribution of arbitrary objects possible.
For Fig.~\ref{fig:means}~(B), a coaxial connector has been placed in between the two telescope arms,
 and the material budget has been reconstructed from the multiple scattering kink angles at the different positions within the material.
The structures of the connector are well-resolved. 
Measurements similar to the one presented here have the potential to be used to produce full tomographic images by rotating the object
 and repeating the measurement for different angles and particle energies~\cite{Tipp2017-Tomo}.

\section{Summary and outlook}

This contribution presents a measurement of the scattering angle distribution of various scatterers
 using charged particle trajectories reconstructed with a precise beam telescope.
This allows for the calculation of the material budget under the assumption of a model, e.g.\ Highland.
The procedure can be calibrated by measuring the material budget of targets of precisely known thickness.
Using this method, the position-resolved material budget of arbitrary targets can be measured in a plane parallel to the sensor planes.
This method could be used to e.g.\ characterise the material budget of future particle detector modules, where a precise knowledge of the material budget is of concern.
Furthermore, the methodology could be extended by recording scattering images of targets from different angles in order to reconstruct tomographic images.

\section*{Acknowledgement}

The measurements leading to these results have been performed at the Test Beam Facility at DESY Hamburg (Germany), a member of the Helmholtz Association (HGF).

{\small
\bibliographystyle{unsrt}
\bibliography{bibliography}
}
\end{document}